\documentclass[11pt]{article}
\newcommand{\beq}{\begin{equation}}
\newcommand{\eeq}{\end{equation}}
\newcommand{\beqs}{\begin{eqnarray}}
\newcommand{\eeqs}{\end{eqnarray}}
\usepackage[pdftex]{hyperref}
\usepackage{txfonts}
\begin{document}

\title{A critique of recent theories of spin half quantum plasmas}

\author{\sc Govind S. Krishnaswami$^{1}$, Rajaram Nityananda$^{2}$, Abhijit Sen$^{3}$ and \\ \sc Anantanarayanan Thyagaraja$^{4}$
\\ \\
$^{1}$Chennai Mathematical Institute,  SIPCOT IT Park, Siruseri 603103, India\\
$^{2}$National Centre for Radio Astrophysics, TIFR, Pune 411007, India\\
$^{3}$Institute for Plasma Research, Bhat, Gandhinagar 382428, India\\
$^{4}$Astrophysics Group, University of Bristol, Bristol, BS8 1TL, UK
\\ \\
Email: {\tt govind@cmi.ac.in, rajaram@ncra.tifr.res.in,} \\ {\tt senabhijit@gmail.com, a.thyagaraja@bristol.ac.uk }}

\date{June 5, 2013}

\maketitle

\begin{abstract}

In this Letter certain fundamental physics issues relating to recent theories of so-called ``spin quantum plasmas'' are examined. It is shown that the derivations and some of the results obtained in these theories contradict well-established principles of quantum mechanics, especially in their treatment of fermions and spin. The analysis presented suggests that the aforementioned theories do not apply for any range of temperatures at the stated densities and furthermore fail to make any experimentally accessible and testable predictions.

\end{abstract}


{\bf PACS:} 52.35.-g, 52.35.Hr; 52.35.We; 67.10.-j; 67.30.hj \\

\vspace{.2cm}

In the last few years a number of publications have appeared in which spin-1/2 quantum plasma equations (sometimes called Spin Quantum Hydro Dynamics [SQHD]) are derived and applied to various problems; typical works are \cite{Marklund,MahajanBraun,MahajanAsenjo} and many papers cited there and in the review\cite{shukla}. The purpose of this Letter is to raise certain fundamental quantum physics issues which appear to have been overlooked in these works. For example, the recent review article\cite{shukla}, quotes identical formulae for the ``spin magnetisation density'' (in Eq.(18), p. 890 and on p. 901 following Eq.(97) {\em et seq.}), citing \cite{Marklund}, in apparent contradiction to its own stated views on the theory of the degenerate electron gas. Unless clarified and related to well-known works in the quantum theory of Fermi systems\cite{LifshitzPitaevski, AshcroftMermin, Kittel}, the results derived in these works would appear to be inconsistent with well-established physical principles. Indeed, critiques of this nature, related to (but not identical with) the point of view adopted here, have already appeared [cf. Vranjes {\em et al}, \cite{Poedts1}, Bonitz {\em et al}, \cite{Bonitz}]. Our work differs from \cite{Poedts1} in its specific focus on the erroneous treatment of fermions in the cited papers on spin quantum plasmas. It also differs from the important critical study by Bonitz {\em et al}\cite{Bonitz} where the authors compare ``quantum hydrodynamical'' (QHD) predictions \cite{shukla2} [though not those associated with intrinsic spin effects, which is the subject of this Letter] with density functional (DFT)-based simulations.

We consider first the paper by Marklund and Brodin\cite{Marklund}, as this appears to form the basis for all subsequent works in this area. The authors make the explicit claim: ``In this Letter, we present for the first time the fully nonlinear governing equations for spin-1/2 quantum electron plasmas. Starting from the Pauli equation describing the non relativistic electrons, we show that the electron-ion plasma equations are subject to spin-related terms. These terms give rise to a multitude of collective effects, of which,  some are investigated in detail. Applications of the governing equations are discussed, and it is shown that under certain circumstances {\em the collective spin effects can dominate the plasma dynamics.}''[our italics]. 

We have two major (related) concerns with the theory of spin quantum plasmas, as presented in the papers cited. Firstly, the derivation of the ``spin quantum plasma equations'' in \cite{Marklund} starts with the independent electron approximation in which the Coulomb interaction of the electron gas is neglected and the single-electron, non-relativistic Pauli equation is invoked [cf. Eq.(1) of \cite{Marklund}]. This assumption was also made by Sommerfeld in his theory of the free-electron gas in a metal \cite{LifshitzPitaevski,AshcroftMermin,Kittel,RASmith,Ziman,KHuang}. However, in the paragraph preceding this equation, the authors state that they assume [unlike Sommerfeld who explicitly invokes Fermi-Dirac(FD) statistics for the electron gas] the simple product representation of the $N$-electron wave function of the system. They state, apparently in justification: ``Thus, we will here neglect the effects of entanglement and focus on the collective properties of the quantum electron plasma''. In our view, it is a serious error to ignore anti-symmetrization of the many-electron wave function and FD statistics. This plainly contradicts Pauli's exclusion principle. The latter is not only a mathematical consequence of relativistic quantum field theory via Pauli's own ``spin and statistics'' theorem \cite{Pauli}, but is also supported by numerous experimental results accumulated over a century on fermions in atomic and condensed matter physics.

The authors' neglect of the special type of electronic correlation  implied by the exclusion principle, especially when dealing with any many-identical-fermion system when the temperature is well below the Fermi temperature [defined for example in \cite{AshcroftMermin,Kittel,RASmith} and in Eq.(\ref{tfermi}) below] vitiates their theory. This neglect is contrary to the fundamental principle of condensed matter theory, that when the electron thermal de Broglie wavelength [$\lambda=\frac{\hbar\sqrt{2\pi}}{(m_{e}T)^{1/2}}$] is of order or larger than the inter electron distance [$n_{e}^{-1/3}$, see \cite{KHuang}, p. 226], the exclusion principle constraints and FD statistics are not dispensable {\bf luxuries but mandatory necessities}. Marklund and Brodin's use of the simple product wave function is only acceptable for {\em distinguishable fermions} or when the Fermi gas is so hot that Maxwell-Boltzmann(MB) statistics applies - as it does in fusion plasmas, for example, where MB distributions describe thermodynamic equilibria and approximate local thermodynamic equilibrium under collisional conditions [cf.\cite{RASmith},p.42]. This can happen only when $\lambda \ll n_e^{-1/3}$. When this condition fails to hold, it is essential to use Slater determinantal wave functions or a second-quantized formalism in working out all {\em average, fluid properties} of the electron gas \cite{LifshitzPitaevski,AshcroftMermin}. The failure to do so can result in some strange properties being assigned to the electron gas, at variance with both standard theory and experiments. It is well-known[{\em op. cit.} \cite{LifshitzPitaevski,AshcroftMermin,Kittel,RASmith,Ziman}] that  the negligible electronic contribution to the specific heat and magnetic properties such as the smallness of Pauli spin-paramagnetism and Landau diamagnetism are direct consequences of FD statistics of the electron gas. 

Secondly, we note that an immediate consequence following from the authors' neglect  of FD statistics for the electron gas is that, their formula for the magnetisation of the electron plasma- treating, as they do, the ion fluid as a uniform neutralising background for simplicity- is grossly in error. According to them, the magnetisation spin current is,
	\beq
	\label{jspin}
	{\bf j}_{\rm sp} = \nabla \times [2n\mu_{B}{\bf S}]
	\eeq
Here, $n$ is the conduction electron number density and ${\bf S}$ is the local average ``spin vector'' of the electrons at the elementary volume over which $n$ is reasonably constant. Marklund and Brodin identify $\mu_{B} B_{0}/T$ as a dimensionless measure of quantum effects. Here $B_0$ is the ambient external magnetic field, $\mu_{B}=\frac{e\hbar}{2m_{e}}=9.4\times10^{-24}$A.m$^{2}$, the Bohr magneton and $T$, the electron temperature (we measure temperatures in energy units and use SI units throughout). The numerator is evidently the intrinsic electron spin-magnetic moment energy in the external field. It is known that the classical thermal Larmor gyro motion of the electron in a magnetic field is associated with a Larmor gyromagnetic moment such that $\mu_{L}B_{0}=E_{\perp}=\frac{2}{3}E \simeq T$. Hence, the parameter in question is simply, $\mu_{B}/\mu_{L}$. Marklund and Brodin correctly state that in high temperature plasmas or in the presence of significant fields, this parameter is small and hence ``spin quantum effects'' are negligible in comparison with the usual gyromagnetic moment effects. What they {\em fail} to mention is the existence and physical significance of the  {\bf degeneracy parameter}, defined by $D=T/T_{\rm F}$, where the Fermi temperature
	\beq
	\label{tfermi}
	T_{\rm F} = \frac{\hbar^2}{2m_e} (3\pi^2 \: n)^{2/3}.
	\eeq
In order to apply MB statistics, it is essential that $D \gg 1$\cite{RASmith}.   This  ``degeneracy condition'' for FD statistics is equivalent to the principle stated earlier; thus, $D\gg 1$ implies, $\lambda \ll n^{-1/3}$.  In metals, with $n \simeq 10^{28}$ to $10^{29}$m$^{-3}$, $T_{\rm F}$ is a few  eV [cf. Table 1.1 and Table 2.1 in \cite{AshcroftMermin}]. Hence at room temperature, the conduction electron gas is highly  degenerate [$D \simeq 10^{-2}$] and the authors' claim that at ``low temperatures'' quantum spin effects could be important is essentially incorrect since the spin  magnetisation ought, by Pauli ``blocking'', to be of order $2(n_{+}-n_{-})\mu_{B}{\bf b}$, where ${\bf b}$ is the unit vector in the local magnetic field direction and $n_{\pm}=\frac{n}{2}[1 \pm \frac{3\mu_{B}B_{0}}{2T_{\rm F}}]$. Here $n_{\pm}$ refer respectively to the number of electrons per unit volume with their intrinsic spin vectors parallel and antiparallel to the local magnetic field. Note that $n_{+}+n_{-}=n; \frac{n_{+}-n_{-}}{n}=O[\frac{T}{T_{\rm F}}]$  [cf. \cite{RASmith,Ziman,KHuang}]. Physically this means that at temperatures ``low'' compared to the Fermi temperature [cf. Eq.(\ref{tfermi})], which depends solely on the electron number density and physical constants, only electrons in a layer close to the Fermi level [estimated by $n(\frac{T}{T_{\rm F}})$] contribute to the magnetisation current\cite{AshcroftMermin,Kittel}. This is also the direct consequence of the kinetic theory of Fermi liquids [cf. \cite{LifshitzPitaevski}] from which any reasonable {\em fluid} theory of the electron plasma ought to be derived using an appropriate Chapman-Enskog asymptotic expansion in powers of the relevant Knudsen number (inverse collisionality parameter measuring the departure from local thermodynamic equilibrium \cite{Ziman,LifshitzPitaevski}). Essentially the same arguments underly the standard theories of electronic specific heats and Pauli spin paramagnetism and Landau diamagnetism. This analysis of the situation shows that the ``average intrinsic spin vector'' ${\bf S}$ appearing in the above equations cannot possibly be of unit magnitude but must be of order $[\frac{n_{+}-n_{-}}{n}]{\bf b}$, where ${\bf b}$ is a unit vector. On the other hand, when $T\gg T_{F}$ and MB statistics apply [as happens in all fusion and astrophysical plasmas except in white dwarf cores], the quantum intrinsic spin effects are entirely negligible both on individual electrons and collectively. Under these conditions, an individual electron possesses a ``Larmor magnetic moment'', $\mu_{\rm L}=T/B$ in a magnetic field of magnitude $B$. The magnetic moment from its intrinsic spin has a magnitude $\mu_{\rm B}= \frac{e\hbar}{2m_{e}}$.  It is clear that the force felt by the electron in an inhomogeneous field is many orders of magnitude larger than that due to its intrinsic spin in any magnetic field since for all cases $\mu_{\rm L} \gg \mu_{\rm B}$, as noticed previously and also noted by Marklund and Brodin. 

In view of the above discussion, it is easy to see that the papers by Mahajan and collaborators\cite{MahajanBraun,MahajanAsenjo} suffer from contradictory assumptions in dealing with the electron gas in a metal. On the one hand they explicitly assume that ${\bf S}$ is a unit vector which leads to a very large estimate for the ``spin magnetisation current density'' given by the Marklund-Brodin formula Eq.(\ref{jspin}) quoted above. Thus, we may estimate, $j_{\rm sp} \simeq  2n\frac{\mu_{B}}{L_{S}}\simeq 10^{5}$ to $10^{6}{\rm A.m}^{-2}$, where, $L_{S}$ is the gradient length-scale of the spin field.  We note that equilibrium conditions require that this should also be of the same order as the density length-scale. The above estimate corresponds to a deliberately chosen ``macroscopic'' scale of 1m. Any smaller choice [e.g. $L_{S} \simeq 10^{-2}$m] will make the current density estimate even larger! This corresponds to virtually a ``saturation magnetic induction'' of $0.1$ to 1 Tesla, a value more typical of core electrons in a ferromagnetic material. On the other hand, the authors require an extremely low temperature $T \simeq 30$K in order to prevent electron-electron and electron-phonon collisions totally damping the electromagnetic wave that moves into their highly inhomogeneous spin density-dominated medium. This low collision rate is due essentially to Pauli blocking which was not taken into account in their equations. The problem is that their formalism and equations are valid neither at high temperatures nor at low temperatures. At high temperatures Larmor gyromagnetic moments will dominate over quantum spin effects and Coulomb collisions imply very short mean-free paths at high densities. At low temperatures the exclusion principle makes the assumption of a unit ${\bf S}$ invalid. Furthermore, nowhere do the authors discuss the basic equilibrium state involving significant spin density gradients. In particular, even if their equations are valid, an experimentalist attempting to verify the predicted instability of the light wave would want to know how one might create such a medium. The assumption by the authors of a saturation magnetisation for the degenerate electron gas comes at the cost of Fermi energy, and hence corresponds to creating a highly excited state of the system.  Also required is the formulation of the Poynting theorem which describes the pumping of the wave at the expense of the spin gradients. The authors' model does make a startling prediction: if one shines light (above the cut-off frequency) at a suitable metal at low temperatures with a suitably prepared internal magnetic field of a very large magnitude (of order 1 Tesla) and spin-density field ${\bf S}$ of large spatial variability, it would  amplify, i.e., lase. Our estimates of the spin current density taking into account the conditions suggest that this result is incorrect. Their model equations do not consider the energy equation [more importantly, the equation of state] of the electron gas they discuss, nor the possible strong damping effects of the neglected electron-electron interactions. It is clear that their equations are not derivable from the Fermi-liquid kinetic/transport equations discussed in the existing literature\cite{LifshitzPitaevski}, since our estimates show that in the relevant limits, the terms involving intrinsic spin are far too small. Indeed, fluid-like models can only be applied when the wave length of any perturbation is significantly large compared with the inter particle distance. A comparison between the pressure gradient terms and the ``Bohm potential'' term [cf. \cite{shukla}, Eqs.(17,27)] reveals that when this condition is satisfied, Bohm potential is negligible and certainly less important than neglected off-diagonal stresses due to interactions. This is demonstrated by the estimates: 
	\beq
	\left\vert -\frac{1}{n_e}\nabla P_F \right\vert \; \simeq \; 
	k \; \delta T_F \; \simeq \; 
	\frac{\pi^{2/3}}{3m_{e}} \hbar^2 n_e^{-1/3} k \; \delta n_{e} 
	\eeq
and the Bohm term $\simeq \frac{\hbar^2}{2m_e}k^3 \left(\frac{\delta n_e}{n_e} \right)$ where $k$ is a typical wave number and $\delta n_{e}$ is the density perturbation. We see that in order for the Bohm term  to be comparable with the usual scalar Fermi pressure, we must have $1/k \simeq n^{-1/3}$, when fluid models will certainly breakdown!

In conclusion, the ``spin quantum plasma theories'' (SQHD) proposed in Refs.\cite{Marklund,MahajanBraun,MahajanAsenjo} and employed apparently without change in \cite{shukla} and many authors cited therein appear to be fundamentally flawed in that some of the underlying assumptions employed in their derivation and some of the consequent results contradict well-established principles of quantum mechanics. Our ``Occam's Razor'' position regarding SQHD models is as follows: a ``reduced'' hydrodynamic model based on free-electron gas ideas [thereby neglecting exchange effects and electron-phonon interactions so important for phenomena such as superconductivity] and deriving quantum spin effects from the Pauli equation invoking FD and MB statistics in appropriate limits cannot contain physics different from those already incorporated in Boltzmann equations modelling Fermi Liquid Theory. The new, startling effects, claimed by the proponents of SQHD, seem contradictory to the simple estimates presented and hence to FLT itself. Thus, in our view, in the absence of an acceptable, physically rigorous, Chapman-Enskog derivation of fluid equations based on the extant, reduced Boltzmann-kinetic (FLT) approaches, the applicability of these theories to experimental situations and the predictions of SQHD appear to be questionable. Indeed, our estimates already suggest that such a rigorous derivation will only lead to small corrections negligible in comparison with interaction effects not taken into account in the free-particle Hamiltonians used in Refs.\cite{Marklund,MahajanBraun,MahajanAsenjo,shukla}. It is noteworthy that so far we have found only a single prediction\cite{MahajanBraun} based on SQHD which could be tested in principle against experiment. We note that other authors \cite{Poedts1,Bonitz} have criticised QHD and have pointed to the dangers of misusing fluid models outside their MB realms of validity. Such models apply only at sufficiently high temperature, to classical, low density gaseous plasmas, close to local thermodynamic equilibrium. They are relevant only to long wavelength phenomena, and cannot generally describe collective kinetic effects like Landau damping/phase-mixing. A discussion of neutron star applications is outside the scope of this work. 

We thank Dr. R. Ganesh for his critical reading of the manuscript and his useful suggestions. We are grateful to numerous colleagues for valuable discussions and suggestions. The work of GSK was supported by a DST Ramanujan Fellowship of the Govt. of India.

\end{document}